\documentclass[useAMS,usenatbib,letter]{mn2e}
\usepackage{psfig} 
\usepackage[dvips]{graphicx}
\usepackage{natbib}
\def\mnras{MNRAS}
\def\apj{ApJ}
\def\aap{A\&A}
\def\apjl{ApJL}
\def\nat{Nature} 
\def\apjs{ApJS}

\def\aj{AJ}

\def\erg{erg~s$^{-1}$~cm$^{-2}$}

\def\xte{{\sl RXTE}}

\def\mr{{MR~2251-178}}

\def\deg{$^{\circ}$}

\def\lesssim{\mathrel{\hbox{\rlap{\hbox{\lower4pt\hbox{$\sim$}}}\hbox{$<$}}}}
\def\gtrsim{\mathrel{\hbox{\rlap{\hbox{\lower4pt\hbox{$\sim$}}}\hbox{$>$}}}}
\title[Correlated X-ray/Optical variability of MR 2251-178]{Correlated X-ray/Optical Variability in the Quasar MR2251-178}  
\author[P. Ar\'evalo et al.]{P. Ar\'evalo$^{1}$\thanks{E-mail: patricia@astro.soton.ac.uk}, P. Uttley$^{1}$, S. Kaspi$^{2,3}$, E. Breedt$^{1}$, P. Lira$^{4}$, I. M. McHardy$^{1}$ \\ 
$^1$School of Physics and Astronomy, University of Southampton, Southampton SO17 1BJ, UK\\
$^2$School of Physics and Astronomy and the Wise Observatory, Raymond and Beverley Sackler Faculty of Exact Sciences,\\ Tel Aviv University, Tel Aviv 69978, Israel\\
$^3$Physics Department, Technion, Haifa 32000, Israel\\
$^4$Departamento de Astronom\'ia, Universidad de Chile, Casilla 36-D, Santiago, Chile\\
}

\begin{document}
\date{Received /Accepted}
\pagerange{\pageref{firstpage}--\pageref{lastpage}} \pubyear{2008}

\maketitle
\label{firstpage}
 
\begin{abstract}
Emission from Active Galactic Nuclei is known to vary strongly over
time over a wide energy band, but the origin of the variability and
especially of the inter-band correlations is still not well
established. Here we present the results of our X-ray and optical
monitoring campaign of the quasar \mr , covering a period of 2.5
years. The X-ray 2--10 keV flux is remarkably well correlated with the
optical B, V and R bands, their fluctuations are almost simultaneous
with a delay consistent with 0 days and not larger than 4 days in
either direction. The amplitude of variations shows an intriguing
behaviour: rapid, large amplitude fluctuations over tens of days in
the X-rays have only small counterparts in the optical bands, while
the long-term trends over hundreds of days are \emph{stronger} in the
B band than in X-rays. We show that simple reprocessing models, where
all the optical variability arises from the variable X-ray heating,
cannot simultaneously explain the discrepant variability amplitudes on
different time-scales and the short delays between X-ray and optical
bands. We interpret the variability and correlations, in the
optically-thick accretion disc plus corona scenario, as the result of
intrinsic accretion rate variations modulating both X-ray and optical
emission, together with reprocessing of X-rays by the accretion disc.
\end{abstract}


\section{Introduction}
Optical continuum variability is a universal property of radio-quiet
AGN which has been studied for several decades, but its origin is
still unclear.  The optical continuum emission almost certainly
originates from the accretion disc \citep{koratkarblaes99}, so it is
natural to assume that disc variability (e.g. through accretion rate
fluctuations) drives the optical variability.  Models where the
variability is {\it intrinsic} to the disc, however, suffer from
several problems.  Firstly, optical variability is seen on time-scales
as short as a day, whereas accretion variability time-scales in
standard accretion discs \citep{shakura} should be long in the
optically-emitting regions, comparable to the viscous time-scale which
can be of the order of a year or more \citep{trevesetal88}.  Secondly,
variations are well-correlated in different optical bands, with only
short (days) delays between bands, in the sense that longer-wavelength
variations lag shorter-wavelength variations
\citep{wandersetal97,collieretal01,cackettetal07}.  Since local disc
temperature $T$ should decrease with radius $R$ as $T\propto
R^{-3/4}$, shorter wavelengths originate from smaller
radii. Therefore, in the case of inward-propagating fluctuations the
long-wavelength variations lead, rather than lag, the short-wavelength
variations. Also, the resulting delays would be large, comparable to
the radial drift time-scale, contrary to the observed day-scale
delays.

To account for the observed optical properties, \citet{kroliketal91}
suggested that the variability is driven by changes of the X-ray
continuum which illuminates and thereby heats the disc, causing
optical continuum variations.  Since the X-ray emission is likely to
be centrally concentrated, blue bands should respond to the X-ray
variations first, followed by red, with short delays corresponding to
the light-travel time delay between the blue and red-emitting parts of
the disc.  The predicted optical time-delay $\tau$ scales with
wavelength $\lambda$ as $\tau \propto \lambda^{4/3}$
\citep{collieretal99}, and the observed dependencies of optical
continuum lags on wavelength are consistent with this prediction
\citep{cackettetal07}.

 A more stringent test of reprocessing models is to directly compare
 X-ray and optical variations, using multi-wavelength monitoring
 campaigns.  To date, a handful of these campaigns have been carried
 out, largely facilitated by the {\it Rossi X-ray Timing Explorer}
 ({\it RXTE}) in conjunction with various ground based observatories.
 The results have been mixed, with some AGN showing good evidence for
 correlated variability, as would be expected from reprocessing models
 (e.g.  NGC~4051, \citealt{petersonetal00,shemmeretal03}; NGC~5548,
 \citealt{uttleyetal03}; Mrk~509, \citealt{marshalletal08}), while
 others show more complicated but possibly correlated behaviour
 (e.g. NGC~7469, \citealt{nandraetal98,nandraetal00}) and one AGN
 shows no apparent correlation at all (NGC~3516,
 \citealt{maozetal02}).

The best case for correlated variability to date is NGC~5548, but
\citet{uttleyetal03} note that it is difficult to reconcile the large
fractional amplitude of optical variability on time-scales of
months-years with reprocessing models, since the X-ray variability
shows smaller-amplitude variations than the optical on these long
time-scales.  One would expect smaller relative variations in optical
since the intrinsic emission due to viscous disc heating would dilute
any variable reprocessed component.  Furthermore, Gaskell (2007) notes
that a simple energetics argument rules out reprocessing as the
dominant source of optical variability in many AGN (including
NGC~5548) which have `big blue bumps' dominating the total luminosity,
significantly exceeding the X-ray luminosity available for
reprocessing.  Detailed reprocessing calculations, produced by
\citet{reprocessing} for the case of NGC~3516 show that simple
reprocessing of X-rays on their own could not produce the observed
optical variability. Also, in some AGN with simultaneous X-ray and
optical monitoring, there is evidence that the optical may {\it lead}
the X-ray variations, which might be expected if at least some of the
optical variability is produced by intrinsic accretion fluctuations
propagating through the disc \citep{shemmeretal03,marshalletal08}.

Clearly, no single model provides a satisfactory explanation for all
the data.  It is possible, however, that some combination of
reprocessing and intrinsic accretion variations may explain the
diverse range of optical/X-ray behaviour which is already observed in
only a small sample of AGN.  The location of the optical emitting
region probably plays a key role in determining the balance of
intrinsic versus reprocessed variability.  It is governed by the disc
temperature, which scales inversely with radius $R$ (in units of the
Gravitational radius $R_g=GM/c^2$) but also scales with black hole
mass $M_{\rm BH}$ and accretion rate (as a fraction of the Eddington
rate) $\dot{m}$ as $T\propto[(\dot{m}/{M}_{\rm BH})R^{-3}]^{1/4}$
\citep{trevesetal88}.  Thus over the 3 decade range in black hole mass
expected for AGN, we expect the radius corresponding to a given
temperature to change by a factor of 10 (at the same fractional
accretion rate), perhaps even more for different accretion rates.
Since disc temperature governs the radius where peak optical emission
is produced, \citet{uttleyetal03} suggested that a diverse range of
optical/X-ray behaviour might result from the range of masses and
accretion rates expected in AGN \citep[see also][]{li}.  For example,
the AGN with the most massive black holes will have cool discs and
very centrally concentrated optical-emitting regions, in terms of
gravitational radii. Therefore, the disc variability time-scales will
be short, comparable with the X-ray variability time-scales, so
intrinsic disc variability as well as reprocessing may contribute
significantly to the optical variations.  In contrast, AGN with
lower-mass black holes will have hotter discs so optical emission will
originate from larger radii, where disc variability time-scales are
very long compared to the corresponding time-scales in the innermost
regions, so that only reprocessing may contribute significantly to the
rapid variability we observe.

To test composite models of optical variability, it is necessary to
carry out combined optical/X-ray monitoring of AGN with widely varying
masses and accretion rates.  To date, these campaigns have focused on
Seyfert galaxies, which typically cover bolometric luminosities
ranging from $10^{43}$ to $10^{45}$ erg~s$^{-1}$.  Here we report the
first optical/X-ray correlation for a more luminous radio-quiet AGN,
the quasar \mr\, which we have monitored simultaneously in X-ray and
optical bands for the past 2.5 years.  We describe the data in
Section~2 and show the cross correlations in Section~3. In Section~4
we explore reprocessing and propagating fluctuations scenarios to
explain the X-ray/optical variability in this source and discuss the
implications of our results in Section~5.

\section{Data}
\mr\ was monitored in the X-ray band using the Rossi X-ray Timing
Explorer \xte\ and in several optical bands with the 1.3m SMARTS
telescope in Chile and the 1m telescope at the Wise Observatory. Below
we give a brief description of the observational campaigns and the
construction of the light curves.

\subsection{X-ray monitoring with \xte}
We have monitored \mr\ in the X-ray band with \xte\, taking 1~ks exposure
snapshot every 4.3 days.  In the present paper, we include X-ray data
from 2004 March 27 to 2008 January 13. Within this period, an
intensive monitoring campaign was carried out with daily observations
for three months, between 2007 August 2 and November 1.

Data were obtained using the \xte\ Proportional Counter Array (PCA),
which is sensitive in the range $\sim2.7$~keV to 60~keV and consists
of 5 Proportional Counter Units (PCUs).  Since PCU~0 has lost its
xenon layer leading to a changed instrumental response and high
background, and PCUs 1, 3 and 4 are regularly switched off for our
observations, we only extracted data from PCU~2. We use standard
good-time interval selection criteria (earth elevation, {\sc
ELV}$>10$\deg\ and source pointing {\sc OFFSET}$<0.02$\deg ).
Background data were created using the combined faint background model
and the SAA history file which are all current as of 2008 March.  We
extracted spectra in the 3--12~keV energy range for each snapshot and
used {\sc xspec} to fit a simple absorbed power law model with the
absorption column fixed at the Galactic value, $N_{\rm
H}=2.8\times10^{20}$ \citep{Lockman} to obtain an estimate of the
2--10 keV flux and its error. The top panel in Fig.~\ref{lcs} shows
the resultant 2--10 keV flux light curve.

\subsection{B and V band monitoring with SMARTS}

We used the ANDICAM instrument mounted on the 1.3m SMARTS telescope in
Cerro Tololo, Chile, to take images of \mr\ in the B and V
filters. The observational sampling was set up to match the \xte\
monitoring, starting from August 2006. Observations were taken every
four days between 2006 August 3 and 2007 December 14, with a daily
sampling period in 2007 from August 2 to November 1.

\begin{figure}
\psfig{file=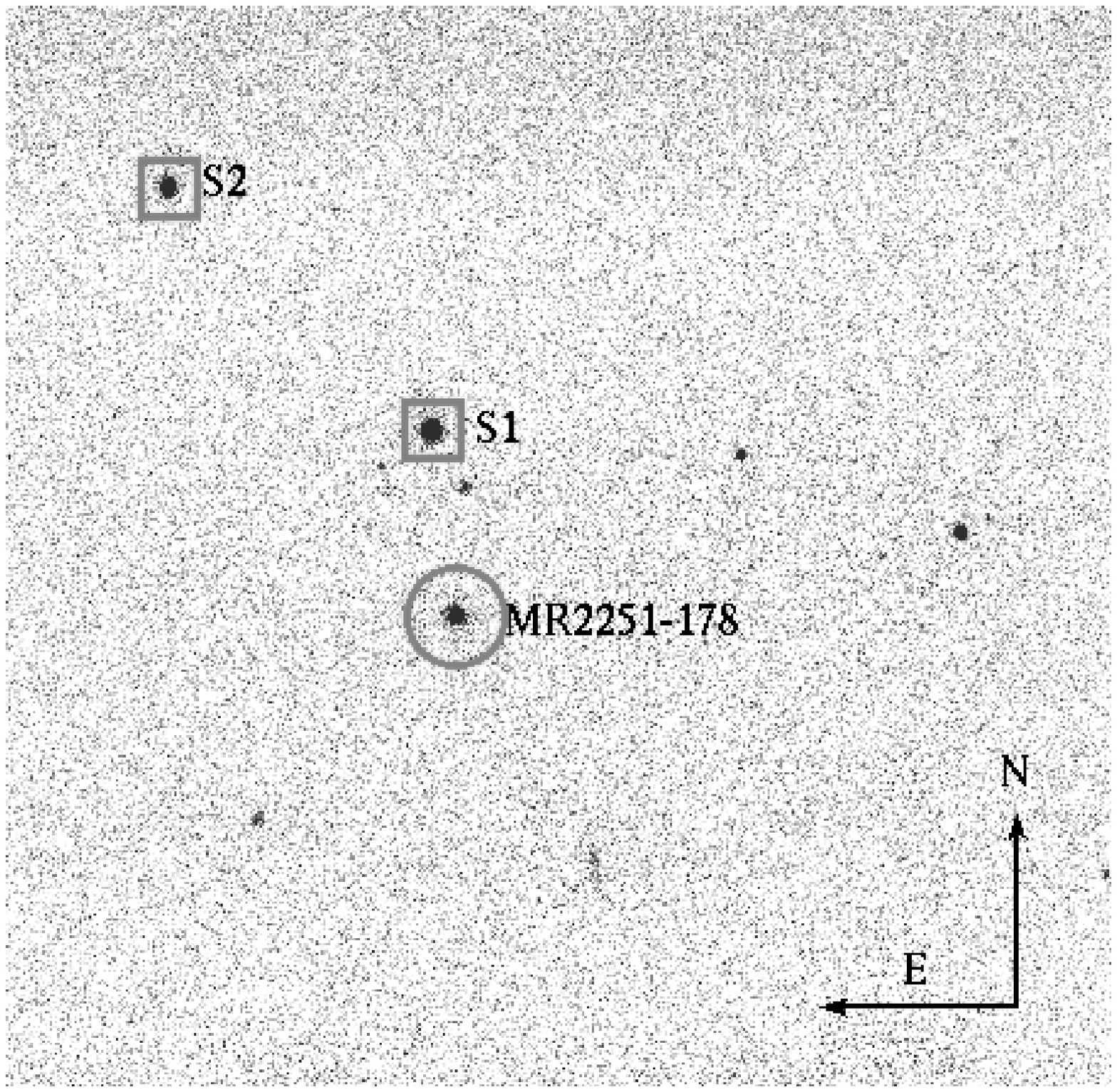,width=8cm}
\caption{SMARTS field of view, the target \mr\ and the two reference stars are labelled. The arrows pointing N and E are 1 arc minute long.}
\label{Simage}
\end{figure}

Two images were taken for each filter and observation date. We
performed aperture photometry on \mr\ and two nearby, non-variable,
reference stars in the field of view, of similar flux to the target,
using an aperture radius of 10 pixels=$3\arcsec.7$. The typical FWHM
of the observations was $1\arcsec.5$. We confirmed that the relative
flux of the reference stars was constant throughout the
campaign. Relative flux light curves were constructed by dividing the
flux of the \mr\ by the combined measurement of both stars. In all
cases where both nightly observations were good, we combined the
relative fluxes, to produce one light curve point per night of
observation. A sample B band image is shown in Fig.~\ref{Simage}, the
target is marked by a circle in the Figure and both reference stars
used are marked by squares and labelled S1 and S2. The source appears
point like in our images and the extraction radius is large enough
that changes in seeing produce a negligible effect in the measured
flux. Using an aperture of 20 pixels produces an identical
relative-flux light curve, with a flux offset of 3\% compared to the
10 pixel aperture radius light curve. Visual inspection of optical
spectroscopic data obtained with the SMARTS 1.5m telescope (not shown
here) does not reveal the presence of any absorption features that
could correspond to the underlying stellar population, so we conclude
that the galaxy contibution to the AGN flux is small.

\subsection{Calibration of the optical lightcurves}
The 1.3m SMARTS telescope observed Landolt standards fields every
photometric night during the campaign to produce zero-point magnitudes
and extinction correction factors for each filter. We used these
zero-points and correction factors, together with the airmass of our
observations, to calculate the magnitude of our reference stars. For
the calibration, we performed aperture photometry on our reference
stars with 20 pixel aperture radius, to match the aperture used on the
Landolt standards, although the resulting magnitudes were almost
identical to the values using 10 pixel apertures. The resulting
average values, using 151 photometric measurements for B and 153 for V
were B=14.00 and V=13.71 for S1 and B=14.90 and V=14.58 for S2, with
errors (rms/number of points) of 0.003 mag. We converted these
magnitudes into fluxes at 5500 \AA\ for the V band and 4400\AA\ for B,
assuming a flat spectrum ($\alpha=0$) within each band. We averaged
the fluxes of both stars and used this value to calibrate the flux
light curves of \mr . The error bars on all optical light curves
correspond to the error on the relative flux and were calculated by
propagating the error on the photometry of \mr\ and the comparison
stars. We do not include the error on the calibration zero-point of
the reference stars because and error in this quantity produces a
shift by the same factor in all the data points and does not
contribute to the poin-to-point scatter in the light curves.

\subsection{B and R band monitoring with the Wise Observatory}

\mr\ was monitored in the B and R optical bands between June 2005 and
December 2007, with observations approximately every 2 weeks up to
December 2006 and then weekly for the rest of the campaign. The data
reduction and light curve extraction were done following the procedure
described in \citet{Wise}. As the field of view of these images is
larger than in the SMARTS data, up to 11 non-variable reference stars
could be used to create a relative flux light curve for \mr . We
scaled the B band relative flux light curve to match the SMARTS B band
data, over the periods of time where the light curves overlapped. To
allow for differences in starlight contamination between the two light
curves, possibly caused by the different apertures and average seeing,
we fitted an additive offset as well as a relative normalisation when
matching the Wise Observatory to the SMARTS fluxes. The small
best-fitting offset, $6\times 10 ^{-13}$ \erg , was added to Wise
Observatory light curve.  For the R band calibration, we obtained the
magnitudes of the reference stars S1 and S2 published in the US Naval
Observatory catalog (USNO V1) to correct the instrumental magnitudes
and to convert these into observed fluxes. The published R magnitudes
are 12.82 for S1 and 13.41 for S2. We used the calibrated flux of
these reference stars to produce the light curve of \mr .

\begin{figure*}
\psfig{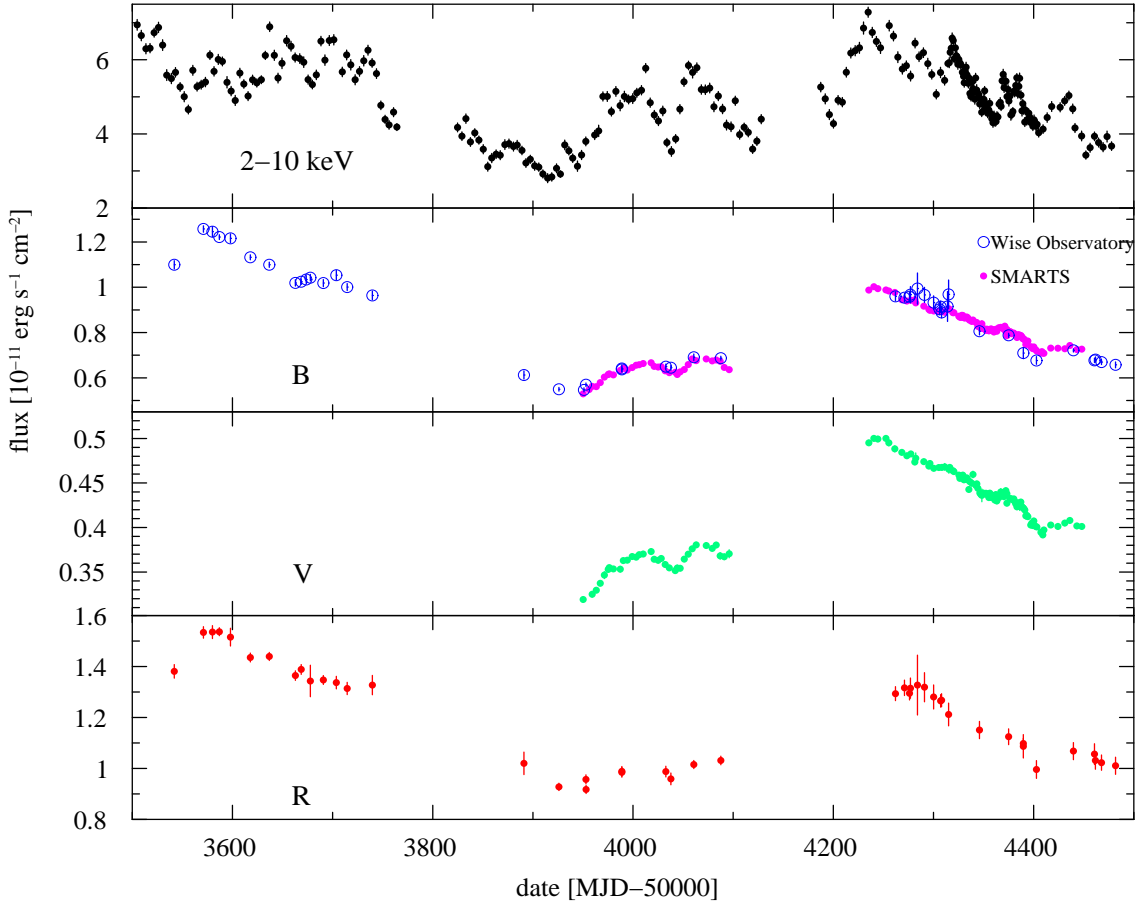}
\caption{\mr\ light curves. From top to bottom: 2--10 keV X-rays; B band Wise Observatory data in blue open circles and B band SMARTS data in pink filled circles; V band SMARTS data, and R band Wise Observatory data. The X-ray light curve is only shown for dates with optical band coverage. Gaps in the light curves correspond to the epochs when the source is not visible from the ground.}
\label{lcs}
\end{figure*}

\section{X-ray/Optical Cross correlation}
As a first step to compare the variability in different energy bands,
we cross correlated the light curves, using the discrete correlation
function (DCF) method of \citet{dcf} and the z-transformed cross
correlation function (ZDCF) as described in \citet{alexander}. The DCF
measures the degree of correlation between the two light curves as a
function of time lag, i.e.\ displacement of one of the light curves on
the time axis. A DCF value of 1 (-1) indicates completely
(anti)correlated data, a value of 0 indicates the data are not
correlated. Unless stated otherwise, a positive value of the time lag
corresponds to X-rays leading.

We first compute the ZDCF between the complete X-ray and B light
curves as this method can deal with sparsely sampled data. The
resulting correlation function has a central peak reaching a maximum
correlation coefficient of 0.71 at a lag of +9.6 days. This central
peak is very broad, with correlation coefficient dropping to 0 at lags
of -142 and +212 days. As is evident from the height of the ZDCF peak
and from the light curves in Fig.~\ref{lcs}, the long term behaviour
of the X-ray and optical bands are well matched, but it is hard to
obtain an accurate lag. To determine the correlation on shorter
time-scales, we split the light curves into year-long segments and
calculated the DCF independently for each segment, avoiding the yearly
gaps. We later combined the resulting DCFs weighting each segment by
the number of DCF points contained. The third segment of the light
curves covers the intensive monitoring campaigns in B, V and X-ray
bands. We re-sampled the intensive light curves taking one point every
four days, to match the observational rate of the rest of the
campaign. This was done to give similar weighting to each
similar-length segment and avoid letting the intensive sampling period
dominate the combined DCF.

\begin{figure}
\psfig{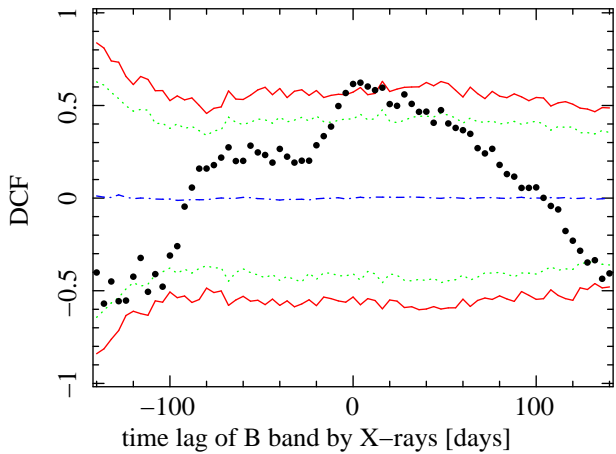}
\caption{The discrete correlation function (DCF) between the X-ray and B bands is plotted in black markers, positive lag values correspond to X-rays leading. The dotted lines represent the 95\% extremes of the distribution of simulated X-ray light curves when correlated with the real B band light curve, the solid lines represent the 99\% extremes of this distribution. The correlation peak around time lag=0 is significant above the 99\% significance level.}
\label{ccf_B_all}
\end{figure}

Figure \ref{ccf_B_all} shows the DCF between the X-ray and B band
light curves, calculated using the three segments shown in
Fig.~\ref{lcs} and including Wise Observatory and SMARTS B band
data. The correlation has a broad central peak reaching a value of
DCF=0.64 (DCF=0.72 for SMARTS data only) and a centroid lag of
+3.6$\pm 9.3$ days. The centroid was calculated as the weighted
average of lags with corresponding DCF values greater than 80 percent
of the maximum, within the central peak. We prefer the centroid to the
DCF peak lag because it is less sensitive to the particular DCF
binning used. The lags and errors were estimated using the random
sample selection method of \citet{peterson}, selecting randomly 67
percent of the data points in the optical light curve together with
their nearest X-ray data point and calculating the DCF centroid for
1000 such trials. The mean lag corresponds to the mean of the 1000
trial centroid-lags and the error corresponds to the rms spread of the
trial centroid-lags.  We repeated the calculation using B band SMARTS
data only and obtained a virtually identical DCF.

The X-ray light curve is equally well correlated with the V band. The
DCF was calculated using the last two segments of the X-ray light
curve and the corresponding segments of the SMARTS V band data, which
was also re-sampled to match the sampling frequency of the rest of the
light curve. The central peak reaches DCF=0.69 with a very similar
shape and lag to the X-ray vs B band case.

The DCF between X-ray and R band data was calculated for the three
light curve segments shown in Fig.~\ref{lcs} and later combined. The
resulting X-ray vs R band DCF has a broad central peak reaching values
of $\sim 0.5$ between lags of -24 and +80 days. The mean centroid lag
is 29 days but the error on the lag is not well constrained. The
significance of all the correlations was estimated using a Monte Carlo
method described below in Sec.~\ref{significance}.

\subsection{Testing the significance of the DCF} 
\label{significance}

We simulated X-ray light curves, uncorrelated with the observed data,
and calculated the DCF between them and the observed optical light
curve. By repeating this process many times, we estimate the
probability of obtaining by chance equal or higher DCF peaks than
observed in the real data. We used the method of \citet{timmer} to
produce red-noise light curves with a given underlying power spectrum
to generate the simulated X-rays. The model power spectrum was chosen
to be a single-bend power law of slope equal to -1 at low frequencies,
bending to a steeper slope at high frequencies, using parameters that
match the shape and normalisation of \mr\ X-ray power spectrum as
measured by Summons et al. (in preparation). Light curve points were
generated in bins of 0.1 days and for a length ten times longer than
the observed X-ray light curve, in order to allow for the long-term
trends which occur in red-noise data. The simulations were then
sampled exactly as the real data and observational noise was added at
the appropriate level. We calculated the DCF between these simulated
light curves and the real optical data in the same way as was done
with the real X-ray data. We repeated this process 2000 times,
recording the DCF values. The mean, 95\% and 99\% extremes of the
distribution of simulated B/X-ray band DCFs are plotted in dot-dashed,
dotted and solid lines, respectively, in Fig.~\ref{ccf_B_all}. As
shown in the figure, the central peak in the real data DCF reaches
higher than 99\% of simulated, uncorrelated light curves. The same is
true for the X-ray vs V band DCF. In the case of X-ray vs R band, the
DCF central peak reached the level of the top 95\% of the simulations,
but did not reach the 99\% level. The R band light curve, however, is
almost identical to the corresponding Wise Observatory B band light
curve. Given that the total, Wise Observatory plus SMARTS, B band
light curve is significantly correlated with the X-rays, it is likely
that the drop in significance in the X-ray vs R correlation is only
due to the smaller number of R data points available.

\subsection{Short time-scale variability}

\begin{figure}
\psfig{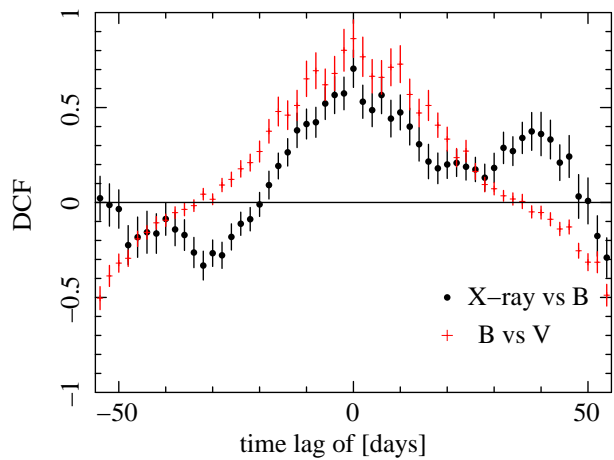}
\caption{DCF calculated using the daily sampled light curves only (MJD 54300 to
 54400) between X-ray and B band in black dots and between B and V bands in red crosses. The X-ray vs V band DCF is very similar to the X-ray vs B band and is not shown for clarity. Positive lag values indicate higher-energy band leading.}
\label{dcf_intensive}
\end{figure}

We used the intensively sampled data taken between MJD 54300 and
 54400, to put tighter constraints on lags between the bands. The DCF
 between X-rays and B band and between B and V bands, of the intensive
 sampling data only, are shown in Fig.~\ref{dcf_intensive}, they both
 show a clear central peak around a lag of zero days. The error bars
 in the Figure are the standard DCF errors described in
 \citet{dcf}. The DCF centroids obtained for the main peaks are $0.6
 \pm 3.1$ days between X-ray and B, $1.3 \pm 3.3$ day for X-ray vs V
 and $0.5 \pm 2.4$ days between B and V. In all cases, positive lag
 values indicate higher-energy band leading. All the lags are
 consistent with 0 days and are constrained to be less than $\sim 4$
 days in either direction. The errors on the lags were again obtained
 using the random sample selection method of \citet{peterson}.

\subsection{Variability amplitudes}
\begin{figure}
\psfig{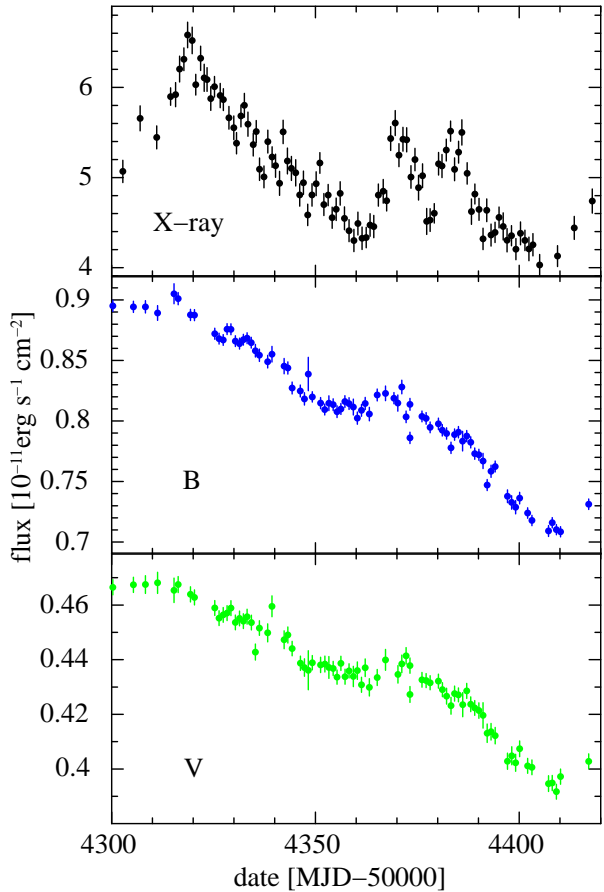}
\caption{X-ray, B and V light curves covering the intensive monitoring campaign. The B and V fluxes have been calculated assuming an effective bandwidth of 968 \AA\ for the B band and 880\AA\ for the V band.}
\label{intensive}
\end{figure}

Figure~\ref{intensive} shows the daily-sampled X-ray, B and V band
light curves. The rapid X-ray fluctuations on time-scales of tens of
days in the X-ray band have only small counterparts in the optical
light curves, with a small or zero lag.  As is clear from this figure
and also Fig.~\ref{lcs}, the long term trends over hundreds of days
appear in all bands. The relative amplitude of these long-term
fluctuations is larger in the B band than in the X-rays: the minimum
to maximum flux ratio observed in the X-rays for the entire period
shown in Fig.~\ref{lcs} is 2.8, between the observations on
MJD-50000=3915 to 4234, while the SMARTS B band data produce a ratio
of 3.14 between days 3950 and 4240. An even larger flux ratio is shown
on the period covered by the Wise Observatory data, when the B band
flux drops by a factor of 4 between 3571 and 3950.

\begin{figure}
\psfig{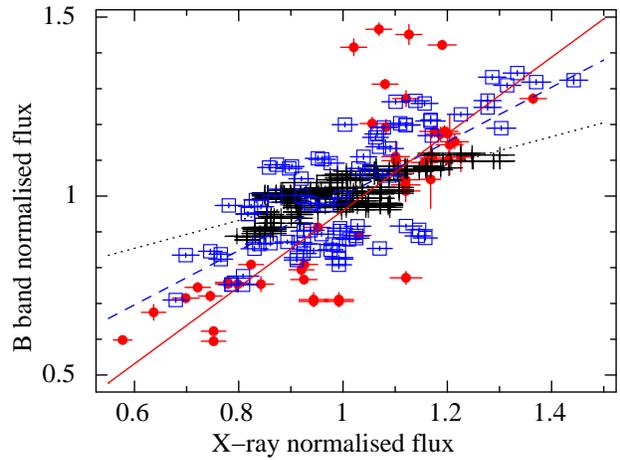}
\caption{Flux-flux plot of the X-ray and B band data on different time-scales, for long (red dots), intermediate (blue squares) and short (black crosses) sampling patterns and light curve lengths. The best fitting linear relation for each data set gets flatter for shorter time-scales probed: $B/\bar B=1.07X/\bar X-0.11$ for the long term light curve (solid lines) $B/\bar B=0.76X/\bar X+0.24$ for the intermediate (dashed line) and $B/\bar B=0.39X/\bar X+0.62$ for the daily sampled, 90 day long light curve (dotted line).}
\label{fluxflux_X_B}
\end{figure}

Figure~\ref{fluxflux_X_B} shows the B band flux as a function of its
nearest (less than 2 days apart) X-ray measurement, each normalised by
the corresponding mean flux of the complete light curve. The dots
correspond to the long term, $\sim$2-week sampled campaign covering
900 days, Wise Observatory data; the squares represent a
shorter, 4-day sampling light curve covering 500 days, and the crosses
represent the intensive monitoring data, sampled daily for 90
days. The best fitting linear relation for each data set gets flatter
for shorter time-scales probed: $B/\bar B=1.07X/\bar X-0.11$ for the
long term light curve (solid lines) $B/\bar B=0.76X/\bar X+0.24$ for
the intermediate (dashed line) and $B/\bar B=0.39X/\bar X+0.62$ for
the daily sampled, 90 day long light curve (dotted line). This
change in slope shows how the B and X-ray light curves, although
always correlated, have different time-scale behaviour, with rapid
fluctuations being much stronger in the X-ray band and long term
trends being stronger in the B band.

\begin{figure}
\psfig{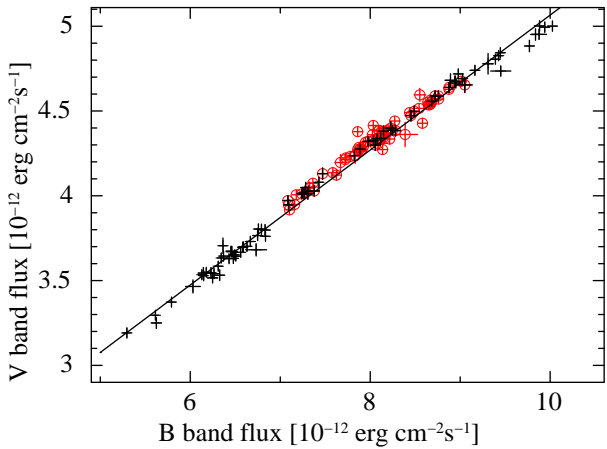}
\caption{Flux-flux plot of the SMARTS V and B band data, the crosses correspond to the 4-day sampling campaign and the open circles represent the intensive monitoring campaign. The flux in both bands varies almost proportionally, following a relation $V=0.4B+1.09\times 10^{-12}$\erg , indicating that the B band varies proportionally more than the V band and that the latter contains a stronger constant offset.}
\label{fluxflux}
\end{figure}

The optical B and V band light curves from SMARTS are almost identical
in shape and only differ in the amplitude of the
fluctuations. Figure~\ref{fluxflux} shows the V flux as a function of
B flux for the same epoch, for the evenly 4-day sampled data and for
the intensive monitoring sample, separately. The flux-flux plots are
almost linear, following a relation $V=(0.398\pm{0.005}) B+(1.09\pm
0.03)\times 10^{-12}$ \erg\ ($\chi^2=137$ for 79 dof) for the
non-intensive segments and $V=(0.375\pm0.01) B+(1.31\pm 0.09)\times
10^{-12}$\erg\ ($\chi^2=93.4$ for 73 dof) for the intensive sampling
data. Notice that the short term light curve has a slightly flatter
flux-flux relation. Normalising each light curve to its mean before
computing the flux-flux relation produces the relative changes of flux
in each band: $V/\bar V= 0.75B/\bar B +0.027$, i.e.\ the relative
amplitude of $V$ fluctuations is smaller than in $B$. This difference
in relative amplitude can be intrinsic to the variability process but
it can also be partly due to starlight contamination producing a
stronger constant component in the V than B band light curves.

\begin{figure}
\psfig{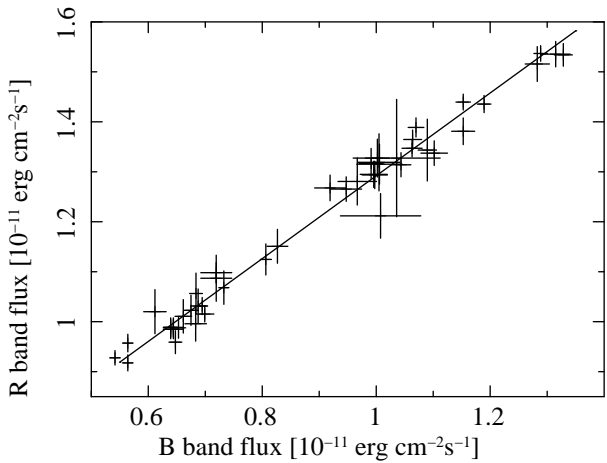}
\caption{Flux-flux plot of the Wise Observatory R and B band data. The flux in both bands varies almost proportionally, following a relation $R=0.83B+4.6\times 10^{-12}$\erg , indicating that the B band also varies proportionally more than the R band and that the latter contains a stronger constant offset.}
\label{fluxflux_R_B}
\end{figure}

The Wise Observatory B and R light curves are also well correlated following a linear relation of $R=0.83B+4.6\times 10^{-12}$\erg\ shown in Fig.~\ref{fluxflux_R_B}. Normalising each light curve by its mean produces a relative variability relation of $R/\bar R=0.62B/\bar B+0.38$. 

\section{The origin of the correlation and lags}
We have shown in the previous sections that the variability in the
X-ray and optical B, V and R bands is well correlated and that any
time lag between optical and X-rays is small ($<4$~days). On
time-scales of months, the fluctuations in the optical bands are
similar to, and even larger than, the X-ray fluctuations while on
shorter time-scales rapid, large amplitude X-ray flares have only
small counterparts in the optical bands. In this section we will
explore possible scenarios to explain the correlated variability
between the bands.

\subsection{Reprocessing}
The small lags between optical and X-ray bands suggest their
variations are connected through reprocessing of X-rays, possibly by
an optically thick accretion disc, where the light travel time to the
reprocessor is the main contributor to the time lags. This scenario
could, in principle, also explain the lack of short time-scale
fluctuations in the optical bands, as these can be smeared out by the
finite light travel time to different parts of the reprocessor.

To test the predictions of this scenario, we generated
thermally-reprocessed light curves, using the observed X-ray light
curves as input. We used the method described in \citet{reprocessing}
to construct the reprocessed light curves. This model assumes a source
of X-rays on the axis of symmetry of the system, illuminating the
accretion disc from a height $h$ in units of gravitational radii
$R_g=GM/c^2$. The optical light curve is constructed by adding the
X-ray flux received at each location in the disc, which depends on the
X-ray source height, to the locally dissipated gravitational energy,
as a function of radius. Each annulus in the disc is then assumed to
emit this total flux as a black body so the characteristic temperature
and flux in a given optical band can be calculated. The light travel
time to the different parts of the accretion disc and from there to
the observer are taken into account to produce the expected delays
between X-ray and optical bands. We left the disc accretion rate $\dot
m$ as a free parameter, which governs the intrinsic, non-variable part
of the optical flux.

The observed X-ray flux was converted to total luminosity assuming a
distance of 271 Mpc to \mr , corresponding to its corrected redshift
of $z=0.063$, using $H_o$ = 73 km/s/Mpc, $\Omega_M=0.27$ and
$\Omega_\Lambda$=0.73. We followed the X-ray spectral fitting of {\sl
  BeppoSAX} data performed by \citet{Orr}, to estimate the total
(0.1-500 keV) X-ray flux incident on the disc and the reprocessed
fraction. We considered a power law model with reflection off cold
material with a small reflection fraction R. A fit with X{\sc spec}
model {{pexriv}}, using $R$=0.16 and $\xi=465.1$ \citep{Orr} produced
a 0.01--500 keV flux of $1.75\times10^{-10}$ \erg , while the 2--10
keV flux was $4.64\times10^{-11}$ \erg. Therefore, we multiplied the
2--10 keV light curve by a factor of four to obtain the total incident
X-rays. We also used the model fit to estimate the fraction of flux
incident on a unit area of the thin disc which is absorbed, and
therefore reprocessed into blackbody photons, obtaing a reprocessed
fraction of 60 percent.

The reprocessed optical light curves were scaled by the same distance
of 271 Mpc to produce observed fluxes. We included an additional
parameter $A$ that multiplies the resulting optical flux to match the
mean observed value. The function of this parameter is to absorb
differences in predicted and observed fluxes due to uncertainties in
the distance and extinction of the observed optical flux.

We varied $h$, $\dot m$, black hole mass, inner truncation radius of
the optically thick disc $r_{\rm in}$, inclination angle to the
observer $\theta$ and rescaling factor $A$ to try to reproduce the
observed B band light curve using the observed X-ray light curve as
input. To first order, the effect of increasing $h$ is to increase the
lag between X-ray and reprocessed optical fluctuations and it also
increases the disc covering fraction. The intrinsic disc accretion
rate $\dot m$ governs the stable disc flux, diluting the variability
of the reprocessed component. Notice that the total optical luminosity
depends on both $\dot m$ and the inner truncation radius of the
disc. The inclination angle introduces smoothing of reprocessed
fluctuations since the response observed on the farther side of the
disc lags the response of the nearer side by a maximum time of twice
the light crossing time of the truncation radius, the maximum being
produced when the disc is viewed edge-on. The black hole mass defines
the size of the gravitational radius, so all the distances and
therefore time-scales scale linearly with this parameter.

\begin{figure}
\psfig{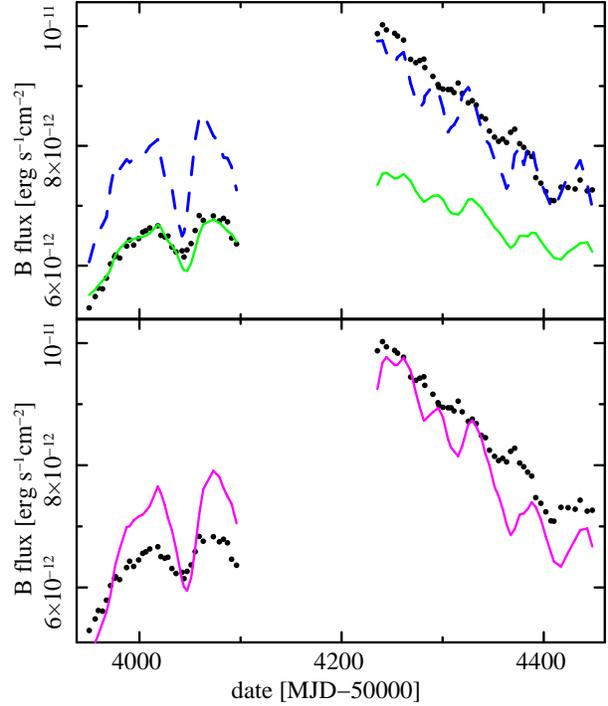}
\caption{\label{reprocess}Reprocessing: The solid and dashed lines represent the reprocessed B band light curves, calculated using the observed X-ray light curve as input. The dots represent the observed B-band light curve. In the top panel the reprocessing model has been fitted to only one segment at the time: the solid line corresponds to the fit to the first segment and the dashed line to the fit to the second. The solid line in the bottom panel shows the best fit to the two segments together.}
\end{figure}

The model light curves corresponding to the best-fitting combinations
of the parameters are shown in Fig.~\ref{reprocess}, together with the
data for comparison. We first fitted each SMARTS B band light curve
segment separately.  The best-fitting parameters for the first segment
shown in Fig.~\ref{reprocess} are $h= 100 R_g, M=10^9M_\odot, \dot
m=0.86, r_{\rm in}=38 R_g, A=0.15$ and for the second segment the
parameters are $h=68 R_g, M=4.5\times 10^8M_\odot, \dot m=0.38, r_{\rm
in}=30 R_g, A=0.33$, both with an edge-on view, $\theta=0.0$. The high
position of the X-ray source seems incompatible with an
evaporating-disc corona but might correspond to the base of a weak
jet. We applied each set of parameters to the other segment to produce
the two pairs of reprocessed light curves, shown by the solid and
dashed lines in the top panel of the figure.  Clearly, the same set of
model parameters cannot simultaneously reproduce the shapes of the two
light curves segments and their relative offsets, which are governed
by long-term trends.  Within a single segment however, the model can
reproduce the short term behaviour moderately well.  This result
immediately suggests that, while the short-term fluctuations may be
caused by reprocessing, the long-term variability should have a
separate origin, e.g. as intrinsic accretion disc fluctuations.

To highlight the difficulty faced by simple reprocessing models in
explaining the long-term optical variability, in the bottom panel of
the figure we show a joint fit to both segments, with resulting
parameters $h= 100 R_g, M=10^9M_\odot, \dot m=0.58, r_{\rm in}=83 R_g,
A=0.46$ and $\theta=0.0 $. The joint fit demonstrates that, in order
to reproduce the long term fluctuations, the model necessarily
produces too much rapid optical variability, even at an edge-on angle
of observation which maximizes the smoothing. Additional smoothing is
produced by having a larger reprocessor i.e. pushing the inner
truncation radius $r_{\rm in}$ outwards, but even then the model cannot smooth
out the rapid fluctuations enough without introducing a noticeable
optical to X-ray lag. Adding more intrinsic, constant, optical flux to
reduce the short-term fluctuations does not solve this issue as this
would also reduce the long term trends, which are at least as strong
in the B band as in X-rays. Therefore, X-ray reprocessing cannot be
the sole cause of all the observed optical variability.

\subsection{Propagating-fluctuations plus reprocessing}

A possible origin for the long-term optical variability is
fluctuations travelling through the accretion disc. In this scenario
the X-ray and optical long term trends can both be modulated by the
accretion rate. A potential problem with this scenario is that if the
accretion rate fluctuations propagate through a geometrically thin
disc, their propagation time can be long and we would expect to see
time lags between the bands, with lower energies \emph{leading}. This
problem can be circumvented if part of the X-rays are reprocessed by
the disc. The large and rapid X-ray fluctuations can then produce some
optical variability where the lower energy fluctuations will
\emph{lag}, therefore canceling part of the propagation effect.

We constructed a propagating-fluctuation model that includes
reprocessing of the X-rays to test whether it can reproduce the
different amplitudes of variability and correct time lags. We followed
the prescription of \citet{lyub}, who proposes accretion rate
fluctuations generated over a wide range in radii in the accretion
disc, on the local viscous time-scale. The fluctuations propagate
inward with viscous velocity through the disc and finally modulate the
X-ray emission at the centre. As the local time-scales decrease with
radius, the more centrally concentrated emission contains the shortest
time-scale fluctuations, seen especially in the X-rays. The resulting
light curves for this model have a $1/f$ power spectrum, bending to
steeper slopes above a characteristic frequency, which is directly
related to the shortest variability time-scales included, i.e., the
viscous time-scale at the inner edge of the accretion flow. In a
standard disc \citep{shakura}, this characteristic time-scale is
proportional to the mass of the black hole and is a function of the
disc thickness ($H/R$) and viscosity ($\alpha$) parameters.

We incorporated reprocessing into the implementation of this model
described in \citet{arevalouttley06}, by adding an X-ray source above
the disc and on the axis of symmetry. We assumed that the optical flux
is emitted thermally by the disc with local emissivity proportional to
the radial gravitational energy release. This emissivity profile is
first modulated by the propagating accretion rate fluctuations and
then also by the variable flux received from the X-ray source.  We
incorporate the travel time of the accretion rate fluctuations to the
central X-ray emitting region and the light travel time of the X-rays
to the different annuli in the accretion disc. For simplicity, only a
face-on viewing angle was considered.

We used the observed X-ray flux and a distance of 271 Mpc as above, to
obtain the X-ray luminosity. The black hole mass of \mr\ has not been
measured but, using the lower limit on the power spectrum break
time-scale of Summons et al.\ (in prep.) and the relation between this
quantity and black hole mass derived in \citet{mchardynat}, we derived
a minimum mass of several times $10^8M_\odot$. We fixed the mass of
the black hole to $8\times 10^8 M_\odot$ and varied disc thickness
and/or $\alpha$ parameter to obtain the correct X-ray power spectrum
from the simulated X-ray light curve, obtaining a value of
$(H/R)^2\alpha=0.08$ at the innermost radius. As, in the model, the
time-scales grow with black hole mass and decrease with $\alpha$,
these parameters are largely degenerate so we only picked one set that
fits the data but note that other pairs would work equally well. We
then fixed these parameters and varied the height of the X-ray source
and intrinsic mean disc accretion rate, to produce different
realisations of the optical light curves.

One possible geometry for the accretion disc/corona system is a
geometrically thin accretion disc that thickens towards the centre to
produce a geometrically-thick, optically-thin corona. As the thickening of the
flow produces shorter time-scale fluctuations at the same radius, the
accretion rate that finally modulates the X-rays can have fluctuations
on time-scales much shorter than the optical emission from the
truncated thin disc. We kept the thickness to radius ratio $H/R$
constant down to a truncation radius of $r_t=10R_g$, with a value of
$(H/R)^2\alpha=0.04$, this produces viscous fluctuations up to
time-scales of 300 days at the truncation radius. Inside $r_t$, we
allowed $H/R$ to grow with decreasing radius as
$(H/R)^2\alpha=0.04(r_t/r)^{1.4}$ to reach the desired value of
$(H/R)^2\alpha=0.08$ at $r=6R_g$ producing fluctuations on time-scales
down to 5 days. We allowed the X-ray source to be higher than the top
of the thick disc, but assumed that it was modulated by the thick flow
fluctuations. The X-ray source height, however, had to be kept small
at 4--6 $R_g$ to produce sufficiently little reprocessing to reproduce
the small relative size of the rapid optical fluctuations.

\begin{figure}
\psfig{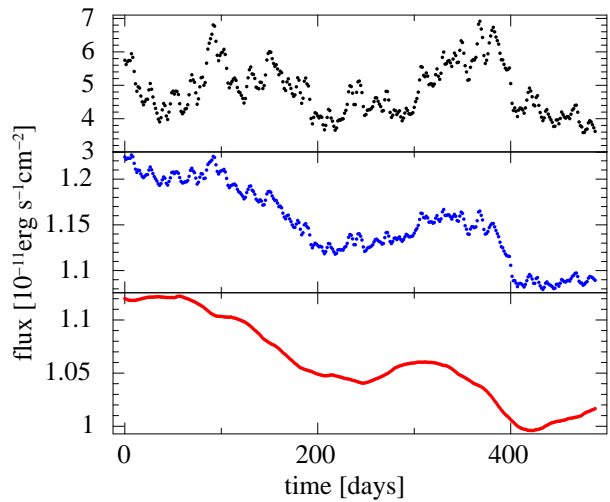}
 \caption{\label{sim_lcs}Simulated light curves of a propagating fluctuation model. The top panel represents the fluctuations in the innermost region, corresponding to the X-ray light curve, the bottom panel shows an accretion-disc emitted B-band light curve, modulated only by accretion rate fluctuations, the middle panel shows this same B band light curve when the effect of reprocessing of X-rays is incorporated.}
\end{figure}

\begin{figure}
\psfig{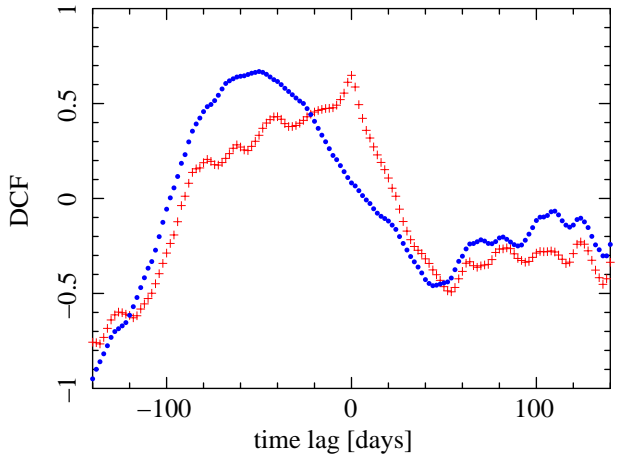}
\caption{\label{dcf_X_B} DCF between the simulated X-ray and B band light curves shown in Fig.~\ref{sim_lcs}, including reprocessing of X-rays (crosses) and not including reprocessing (dots). The additional B band fluctuations produced by reprocessing of X-rays move the DCF peak to 0 lag.}
\end{figure}

\begin{figure}
\psfig{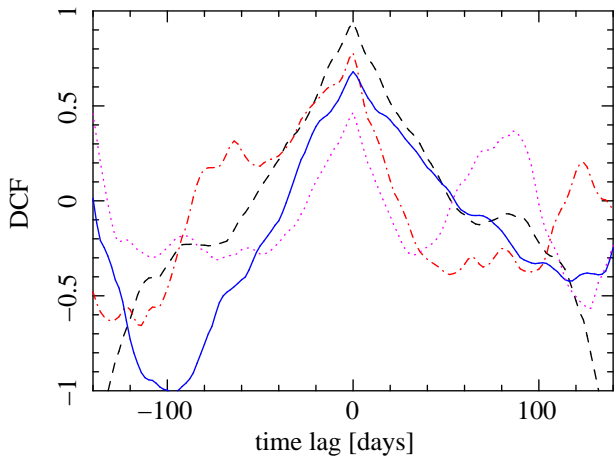}
\caption{\label{dcf_X_B_multi}Sample of DCFs between different realisations of simulated X-ray and B band light curves, with the same input parameters and length as those shown in Fig.~\ref{sim_lcs}. Here we plot the DCFs as continuous lines for clarity. The main DCF peak appears consistently around 0 lag but the height of the peak and the direction of its asymmetry depend on the realisation.}
\end{figure}

Figure~\ref{sim_lcs} shows 500 day long simulated light curves. The
top panel represents the fluctuations in the innermost region,
corresponding to the X-ray light curve, the bottom panel shows a B
band light curve modulated only by accretion rate fluctuations and the
middle panel shows this same B light curve when the effect of
reprocessing of X-rays is incorporated. As expected, the resulting
X-ray and optical light curves are well
correlated. Figure~\ref{dcf_X_B} demonstrates the effects of
reprocessing on the DCF, the dots show the DCF between the simulated
X-ray and B band light curves if reprocessing is switched off
(i.e. between the light curves shown in the top and bottom panels in
Fig.~\ref{sim_lcs}), the B band leads the X-rays by approximately 50
days, which in this case is the travel time of the accretion
fluctuations from the main B emitting region to the centre. If
reprocessing is switched on, so that the rapid X-ray flares are
imprinted as small optical fluctuations in the same light curve used
above (shown in the middle panel of Fig.~\ref{sim_lcs}), the DCF peak
shifts to a lag of zero days, represented by the crosses in
Fig.~\ref{dcf_X_B}. The exact shape of the DCF between B and X-ray
light curves changes with different realisations of the simulation
even when exactly the same setup and parameters are used. In
Fig.~\ref{dcf_X_B_multi} we show a few examples of such DCFs,
calculated for different light curve segments, each 500 days
long. Given the changes in DCF peak height and direction of asymmetry,
we cannot draw any conclusions from these structures in either the
simulated or observed DCFs.

The amplitude of short term fluctuations is strongly reduced in the
optical compared to the X-rays and higher energy optical bands display
larger variability than lower energy ones. The flux-flux relation of
the simulated B and V light curves is similar to the real data,
following a linear relation of the form $V=0.46B+1.9\times 10^{-12}$
\erg . In the model, the temperature radial profile of the accretion
disc produces an emitting region that is more extended in the V than
in the B band, therefore diluting more strongly in the V band
variability. This causes the lower variability amplitude in V,
especially on short time-scales, without including any starlight
contribution that might dilute the variability further.

\begin{figure}
\psfig{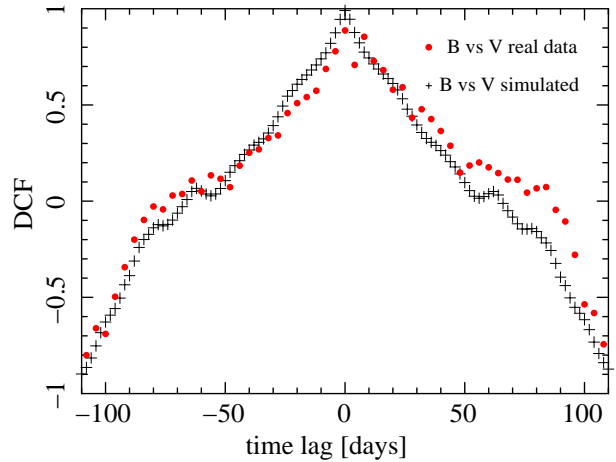}
\caption{\label{acf} DCF between B and V band light curves for the real data (dots) and propagating-fluctuation plus reprocessing simulated data (crosses).}
\end{figure}

The auto and cross correlation functions between optical light curves
can also be reproduced. Figure~\ref{acf} shows the similarity of the
DCFs between B and V bands of the real data and the simulated light
curves shown in Fig.~\ref{sim_lcs}. These functions are almost
symmetrical and very similar to the ACF, so we only display the DCFs
for clarity. The simulations for this particular realisation reproduce
the right time-scale dependence of the variability and produce a lag
between B and V light curves consistent with 0.

A feature which this simple implementation fails to reproduce is the larger amplitude of long term optical fluctuations compared to the X-rays. This effect can appear for example if the fluctuations are damped as they propagate inward or as they transfer from the optically thick flow to the corona. As we only attempted a phenomenological approach to this fluctuating accretion model, we can easily introduce a damping parameter that reduces the X-ray long term fluctuations but this would not produce additional information on the physical processes involved.

\section{conclusions}

We have carried out a simultaneous monitoring campaign in X-ray and
optical B, V and R bands on the quasar \mr\ over two and a half
years. All bands are significantly variable and their fluctuations are
well correlated. The cross correlation functions show peaks around a
lag of zero days, significant over the 99\% confidence level when
compared to uncorrelated simulated data. All the delays between the
bands are consistent with 0 days. We used daily sampled light curves
in X-ray, B and V to constrain the lags between these bands, obtaining
values of $0.6 \pm 3.1$ days between X-ray and B, $1.3 \pm 3.3$ day
for X-ray vs V and $0.5 \pm 2.4$ days between B and V lags, where
positive values indicate higher energy band leading. The lag between
X-ray and R band was determined using the long but more sparsely
sampled light curve, obtaining a value of -4.5$\pm$16.8 days.

The long term trends, over hundreds of days, are well matched by all
the bands observed and are stronger in the B band than in X-rays,
while the other optical bands show smaller-amplitude trends. The
optical light curves are not corrected by host galaxy contribution,
however, so the decreased amplitude of variability in lower energy
bands might be partly due to a constant galactic contribution. On
short time-scales, of tens of days, there are large X-ray flares which
appear strongly attenuated in the optical bands. The intensively
sampled light curves show that the rapid X-ray flares do have optical
counterparts, but that these are very weak.

Pure reprocessing of X-rays cannot produce both the short and long
time-scale optical variability: if the long term optical fluctuations
were produced by reprocessing, we would expect similarly large rapid
optical fluctuations, unless these could be smoothed by light travel
effects. Any geometry of the reprocessor, however, cannot smooth out
the fluctuations on time-scales of 20--50 days without introducing
time-lags of a similar length, which are not observed. A similar
conclusion was reached by \citet{reprocessing} for the Seyfert galaxy
NGC~3516, where the X-ray-UV variability could not be explained by
simple reprocessing. We show that reprocessing of the observed X-rays
cannot reproduce the optical variability in \mr, either producing
rapid fluctuations that were too large or long term trends that were
too small. The short term fluctuations however could be reproduced
reasonably well with the model if we limited the light curve lengths
to $\sim 100$ days, which reduces the contribution of long term trends
to each light curve segment. Therefore, a moderate amount of
reprocessing acting on an already long-term variable optical emission
can reproduce the observed behaviour.

A simple way to decouple the amplitude of short and long term
fluctuations is to have two distinct processes producing the optical
variability, e.g.\ accretion rate fluctuations plus
reprocessing. The covering fraction of the disc, together with the
ratio of intrinsic disc emission to X-ray heating determine the
fraction of optical flux arising from reprocessed X-rays. Therefore,
the size of the rapid optical fluctuations depends on these parameters
and can be largely independent of the amplitude of the long term
trends. To test this possibility, we generated simulated light curves
and compared their statistical properties with the observed data. In
the model, fluctuations propagate inward through the accretion flow,
modulating the optical emission as they travel through the thermally
emitting region and finally modulate the X-ray emission, assumed to be
located in the centre. This process leads to the long term correlated
variability. As the fluctuations are produced at every radius in the
flow, on a radially dependent time-scale, the centrally emitted X-rays
contain short time-scale fluctuations which are originally absent in
the optical light curves.  The rapid fluctuations, only present in the
X-ray emitting region are then introduced into the optical light
curves through thermal reprocessing.

Lastly, although propagation times across the accretion disc can be
long and produce a large lag between optical and X-ray
fluctuations, reprocessing reduces the lag and can even revert it to a
value of the order of the light crossing time to the reprocessor. For
a black hole mass of $10^9 M_\odot$ the light crossing time of one
gravitational radius is 5000 s, so even for a large source height and
inner truncation radius of 50 $R_g$ the crossing
time would be 4 days, less for a less massive black hole, which is
within the observed limits on the lag. We show using the simulated
light curves, that small-amplitude rapid fluctuations in the optical
band can shift the CCF peak towards a lag of zero days, even when the
long term trends of the optical lead the X-rays by $\sim$50 days.  As
inward propagation of the fluctuations and reprocessing have opposite
effects on the time lag, the relative amount of optical/UV variability
produced by each process determines the sign of the lag. This might be
the case in other monitored AGN where the optical/UV can either lead or lag
the X-rays.

An interesting feature of our fits to the optical variability is that
we expect the disc to subtend a small solid angle with respect to the
X-ray source, in order to reduce the effects of X-ray reprocessing and
produce the correct amplitude of optical rapid fluctuations.  This
small solid angle is consistent with the small amount of reflection
inferred from the broadband X-ray spectrum of MR~2251-178 observed by
{\sl BeppoSAX} \citep{Orr}, which shows only a weak iron line with an
equivalent width of $EW\sim 70$~eV, and modest reflection continuum
(reflected fraction $<0.4$).  It is therefore possible that
MR~2251-178 represents a source in a state where the thin disc is
slightly truncated and/or the X-ray emitting region is at a small
height over the disc.  Optical/X-ray studies of other AGN should
provide further useful constraints on their disc/corona geometries.

\section*{Acknowledgments}
This work has made use of data obtained with \xte , the SMARTS Consortium and the Wise Observatory telescopes. We wish to thank the SMARTS team for the good quality data obtained through their service mode observations. PA and IMcH acknowledge support from STFC rolling grant PP/D001013/1 and PU from an STFC Advanced Fellowship. EB acknowledges support from a Stobie-SALT Scholarship, from the South African NRF and the University of Southampton. PL acknowledges support from FONDECYT grant 1080603.

\label{lastpage}

\end{document}